\documentclass[conference]{IEEEtran}
\IEEEoverridecommandlockouts

\usepackage{cite}
\usepackage{amsmath,amssymb,amsfonts}
\usepackage{algorithmic,graphicx}
\usepackage{graphicx,booktabs}
\usepackage{textcomp}
\usepackage{xcolor}
\def\BibTeX{{\rm B\kern-.05em{\sc i\kern-.025em b}\kern-.08em
    T\kern-.1667em\lower.7ex\hbox{E}\kern-.125emX}}
\begin{document}

\title{AgriLens: Semantic Retrieval in Agricultural Texts Using Topic Modeling and Language Models}
\author{
\IEEEauthorblockN{
Heba Shakeel\IEEEauthorrefmark{1}, 
Tanvir Ahmad\IEEEauthorrefmark{2}, 
Tanya Liyaqat\IEEEauthorrefmark{3}, 
and Chandni Saxena\IEEEauthorrefmark{4}
}
\IEEEauthorblockA{\IEEEauthorrefmark{1}\IEEEauthorrefmark{2}Dept. of Computer Engineering, Jamia Millia Islamia, New Delhi, India}
\IEEEauthorblockA{\IEEEauthorrefmark{3}Dept. of Computer Science \& Applications, Sharda University, Greater Noida, Uttar Pradesh, India}
\IEEEauthorblockA{\IEEEauthorrefmark{4}The Chinese University of Hong Kong, Hong Kong SAR, China}
\IEEEauthorblockA{
Email: heba1907468@st.jmi.ac.in, tahmad2@jmi.ac.in, tanyaliyaqat791@gmail.com, chandnisaxena@cuhk.edu.hk
}
}
\maketitle

\begin{abstract}
 As the volume of unstructured text continues to grow across domains, there is an urgent need for scalable methods that enable interpretable organization, summarization, and retrieval of information. This work presents a unified framework for interpretable topic modeling, zero-shot topic labeling, and topic-guided semantic retrieval over large agricultural text corpora. Leveraging BERTopic, we extract semantically coherent topics. Each topic is converted into a structured prompt, enabling a language model to generate meaningful topic labels and summaries in a zero-shot manner. Querying and document exploration are supported via dense embeddings and vector search, while a dedicated evaluation module assesses topical coherence and bias. This framework supports scalable and interpretable information access in specialized domains where labeled data is limited.
\end{abstract}

\section{Introduction}\label{Tab:intro}
In recent years, the volume of textual information has exploded across domains, with much of it remaining unstructured and difficult to analyze at scale~\cite{adnan2019limitations}. This problem is especially acute in the agricultural sector~\cite{cravero2021use,nismi2023review,chamorro2024systematic}, where valuable insights are buried in news articles, research bulletins, government advisories, and even social media discussions~\cite{agritechtomorrowAgTechTrends,pallotta2024discovering}. Despite this richness, agricultural text data remains underutilized due to the complexity of extracting structured knowledge from diverse and evolving sources.

While other fields such as biomedicine\footnote{https://github.com/BaderLab/Biomedical-Corpora} and finance\footnote{https://github.com/The-FinAI/The-FinData} have made significant strides in mining domain-specific corpora, agriculture lags behind—partly due to limited open-access datasets and the absence of tools tailored for interpreting unstructured agricultural narratives~\cite{nismi2023review}. Recent initiatives, such as the release of the ENAGRINEWS corpus~\cite{drury2023enagrinews}, mark a promising shift by assembling large-scale, publicly available agricultural text collections. However, there remains a pressing need for systems that can derive actionable knowledge from such data and present it in a form accessible to domain experts, researchers, and decision-makers.
\begin{figure}[t]
  \centering
\includegraphics[width=0.99\linewidth]{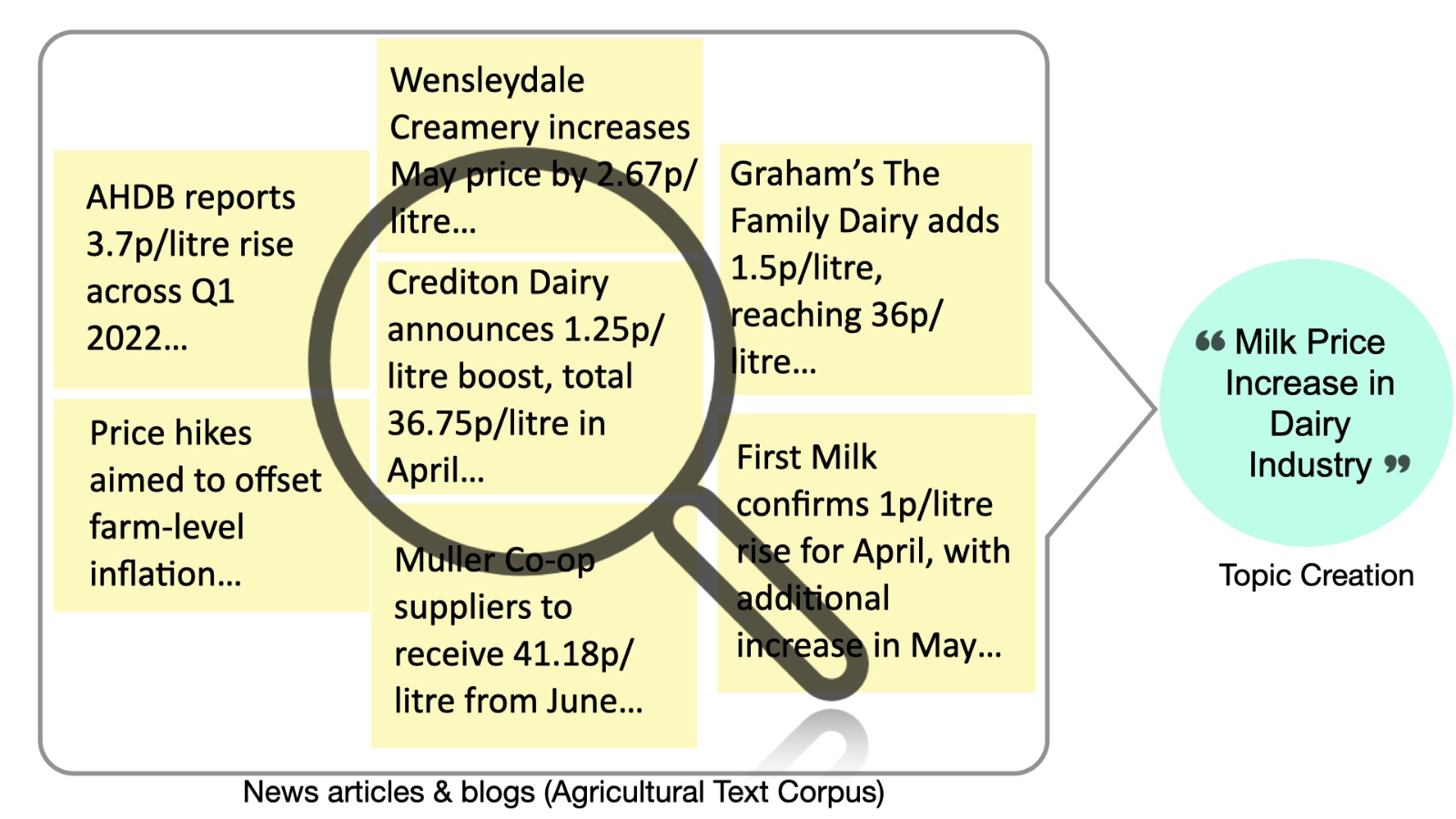}
  \caption{Overview of the AgriLens system: Representation of the topic-level indexing framework. A semantic "Lens" scans a large corpus of unstructured documents to extract and assign concise topic labels.}
  \label{fig:agrilens_system}
\end{figure}
To address these challenges, we introduce AgriLens (Figure~\ref{fig:agrilens_system}), a modular framework that combines fine-grained topic modeling, LLM-based topic labeling, and similarity-based semantic retrieval to help users explore and interact with large-scale agricultural text corpora. AgriLens is designed to be fully unsupervised—leveraging pre-trained language models and embedding-based clustering to extract latent themes without the need for annotated data. It enhances interpretability by generating human-readable topic labels using language models, and enables targeted access to documents via vector search over semantically enriched text chunks.

Our system serves a wide range of users—from agricultural researchers and analysts to development practitioners and information officers—by offering a structured lens into complex and evolving agri-narratives. While AgriLens incorporates core principles of Retrieval-Augmented Generation (RAG), its current implementation integrates semantic retrieval and generation components to support interpretable, topic-driven information access. This approach offers several important advantages. First, it is fully unsupervised, relying on pre-trained language models and embedding techniques—making it well-suited for domains like agriculture, where labeled data is often scarce or unavailable. Second, by automating the topic labeling process using large language models, the system significantly reduces manual overhead while improving interpretability. Users can navigate complex thematic structures without requiring deep NLP expertise. To encourage reproducibility,  we make our code publicly available at GitHub\footnote{https://github.com/HebaShakeel/AgriLens}.
\section{Related Work}\label{Tab:related_work}
 We organize related work into three categories: Topic modeling, RAG systems, and existing semantic search in agricultural data.\\
 \textbf{Topic modeling in agricultural corpora.} Topic modeling has been widely applied across agricultural corpora, from media reports and news articles to research publications. For instance, LDA (Latent Dirichlet Allocation) has been used to track evolving discourse in urban agriculture, uncover COVID-19-related themes in Bangladeshi media~\cite{khatun2021content}, and even correlate disease-related topics with market outcomes like pork price fluctuations~\cite{chuluunsaikhan2020incorporating}. In academic literature, topic modeling helps map thematic structures, such as  \cite{landschoot2024cereal} uses LDA on inter-cropping research to extract $150$ micro-topics. However, traditional models like LDA and NMF (Non-negative Matrix Factorization) often struggle with short or noisy text, lack semantic understanding, and require manual topic labeling. Recent advances such as BERTopic~\cite{grootendorst2022bertopic} address these challenges by leveraging contextual embeddings and clustering, enabling more coherent, fine-grained, and interpretable topic discovery, especially valuable in complex, domain-specific corpora like agriculture~\cite{lin2024research}.\\
 \textbf{RAG systems.}While RAG systems have seen increasing adoption across domains, their integration with topic modeling remains limited. Some studies have explored hybrid pipelines where topic models inform retrieval strategies before generation. For instance, in the education domain, BERTopic combined with LLaMA has been used to extract student-relevant themes for guided question generation and retrieval~\cite{garcia2025llm+}. In agriculture, ShizishanGPT~\cite{yang2024shizishangpt} leverages a domain-specific knowledge base and vector search to support question-answering, demonstrating the promise of embedding-based retrieval tailored to farming queries. AgroLLM~\cite{samuel2025agrollm} similarly applies semantic search via FAISS\footnote{https://faiss.ai} within a retrieval-driven QA pipeline. However, these systems typically emphasize supervised or curated knowledge bases and lack explicit integration of unsupervised topic structures. Our work departs from full RAG frameworks by using topic modeling—via BERTopic~\cite{grootendorst2022bertopic}, as a guiding layer for organizing retrieval.\\
 \textbf{Semantic search in agricultural data.} Semantic search in agriculture has shown advantages over keyword-based retrieval. Beneventano et al.~\cite{beneventano2016exploiting} used AGROVOC\footnote{https://www.fao.org/agrovoc/} and linked vocabularies to enhance retrieval precision, while Kawtrakul~\cite{kawtrakul2012ontology} highlighted early benefits of ontology-driven search in agricultural information systems.
A recent work has explored knowledge graphs and multilingual tools~\cite{peng2023embedding}, but embedding-based semantic retrieval with LLMs remains underexplored. Build on these insights, we propose to develop an unsupervised, topic-driven retrieval framework tailored for agricultural text.

\section{Framework and System Architecture 
}\label{Tab:framework}
\begin{figure*}[t]
\centering
  \includegraphics[width=0.9\textwidth]{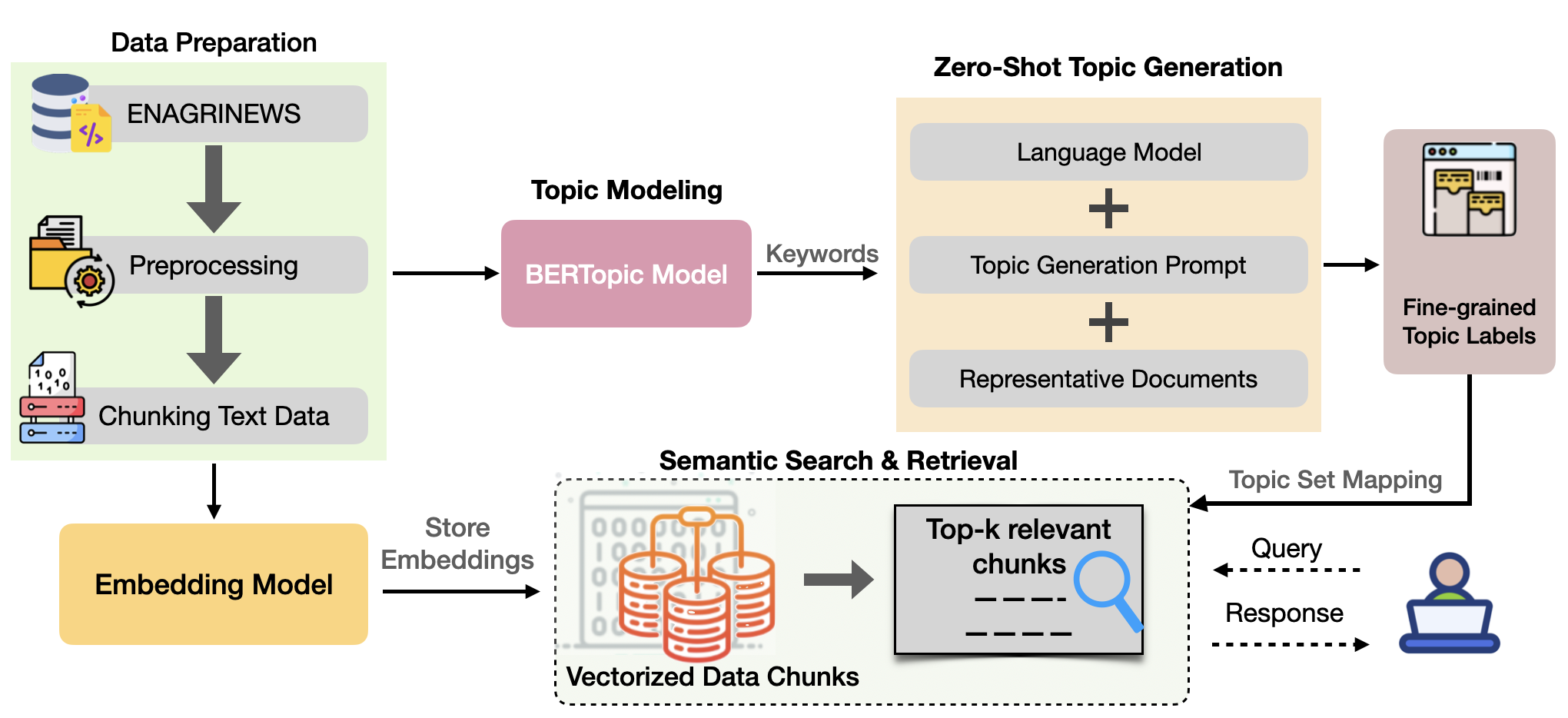}
  \caption{Proposed pipeline for end-to-end semantic search}
  \label{fig:framework}
\end{figure*}

We propose an end-to-end pipeline (Figure~\ref{fig:framework}) for fine-grained topic discovery and semantic retrieval in low-resource domains, using agricultural text from the ENAGRINEWS corpus~\cite{drury2023enagrinews}. The system comprises four modules: Data Preparation, Topic Modeling, Zero-Shot Topic Generation, and Semantic Search \& Retrieval. The subsequent sections provide a detailed description of each of these modules. 
The pipeline starts with the agricultural text corpus, a large collection of agricultural news articles, which is preprocessed and segmented into smaller text chunks for topic modeling and semantic search. BERTopic is then applied to generate topic clusters using transformer embeddings and class-based TF-IDF, producing topic keywords and representative documents. A zero-shot topic generation module refines these into human-readable labels via a prompting-based language model. These fine-grained labels enhance interpretability and are used for topic set mapping in the retrieval stage. The next section outlines the preprocessing techniques used to prepare the ENAGRINEWS data for topic modeling and retrieval.


\section{Data and Pre-processing}\label{Tab:data} This section introduces the dataset~\cite{drury2023enagrinews} and describes the pre-processing steps—extraction, cleaning, and chunking—for vector database storage.
\subsection{Corpus Description}\label{Tab:description}
\begin{figure}[b]
  \includegraphics[width=0.45\textwidth]{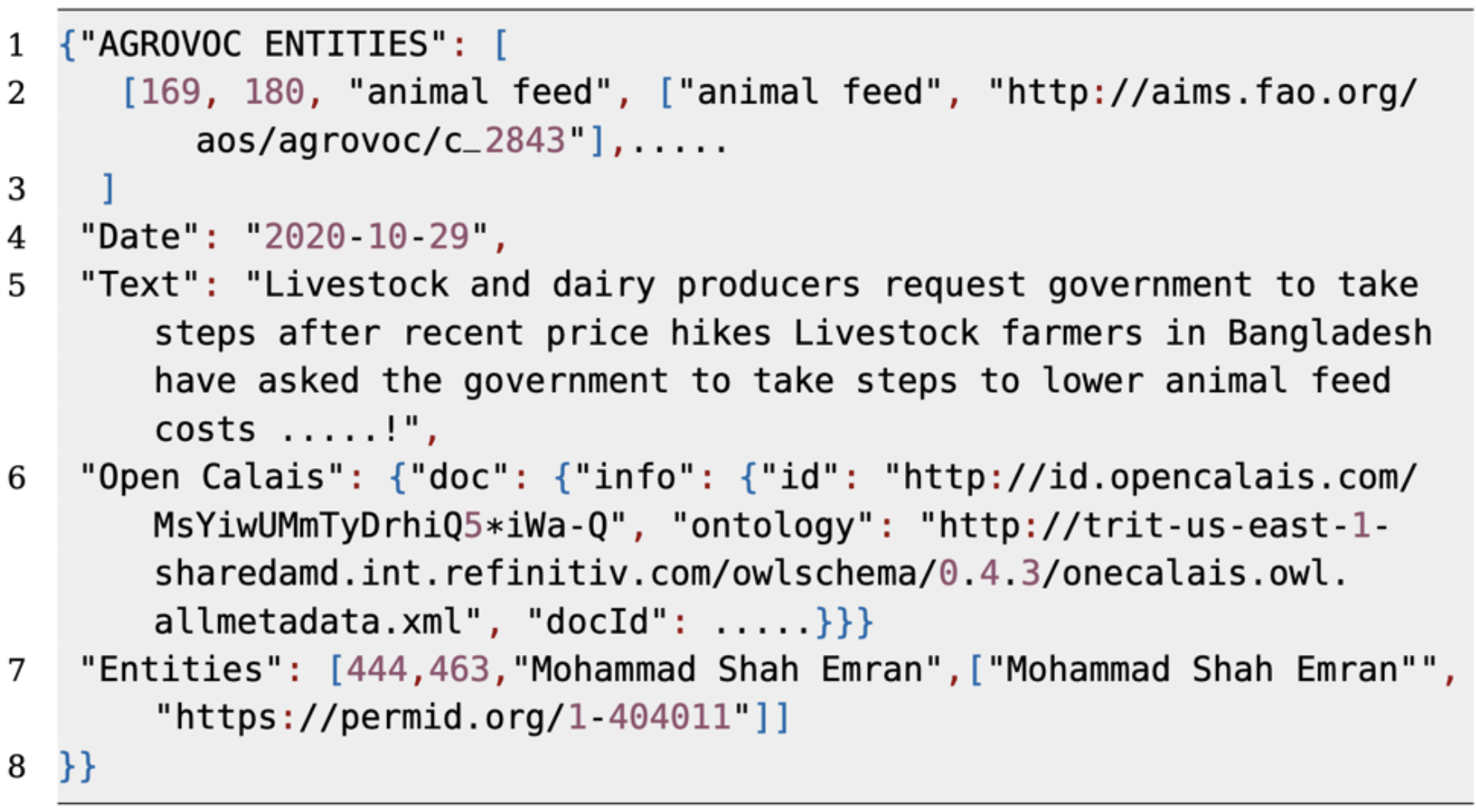}
  \caption{Example snippet from ENARGINEWS document}
  \label{fig:data}
\end{figure}

The proposed approach utilizes the ENAGRINEWS dataset~\cite{drury2023enagrinews}, a large-scale English-language corpus introduced to advance agricultural text mining. Developed in response to the notable gap between agriculture and other domains like biomedicine and finance in adopting NLP techniques, ENAGRINEWS (Figure.~\ref{fig:data}) addresses a key barrier in the field: the scarcity of high-quality, openly accessible text corpora. The dataset contains $22,814$ agriculture-related documents collected from a range of news websites and agricultural blogs over a three-year period (2020–2022). It captures a wide variety of narratives, from market updates and policy discussions to weather impacts and technological trends, offering a diverse and realistic textual landscape for downstream tasks. 
With its combination of scale, domain diversity, and embedded metadata, the data serves as a strong foundation for developing interpretable, domain-specific NLP applications in agriculture.  
\subsection{Text Extraction and Data Cleaning}
To build a clean corpus for topic modeling and retrieval, we automated the traversal of all JSON files in the dataset directory, extracting content from the "Text" field. Files missing this field or containing empty text were excluded to ensure data integrity. Extracted texts then underwent standard cleaning: lowercasing, removing extra whitespace, special and non-UTF characters, and normalizing punctuation. These pre-processing steps normalized the data, reduced vocabulary noise, and enhanced the effectiveness of embedding and topic modeling.
\subsection{Document Chunking}
 Document chunking serves as a critical pre-processing step in the retrieval pipeline of embedding-based systems, especially within frameworks such as RAG. Studies highlight that chunking enables semantic independence, reduces resource usage, and improves retrieval accuracy~\cite{qu2024semantic,ren2025large}. For instance, RAG-based frameworks highlight that chunking long documents into semantically coherent segments improves alignment with query granularity, thereby enhancing retrieval efficiency and the quality of learned embeddings~\cite{wu2024retrieval}.
We adopted a sentence-based chunking strategy: each document is split into sentences using rule-based~\cite{zhao2025moc} sentence boundary detection, then grouped into chunks until a predefined word limit is reached. This approach strikes a balance between coherence and size. In line with recent evaluations of chunking methods, where fixed-size and semantic chunking strategies were compared, we chose sentence-based grouping to preserve contextual flow while managing embedding costs~\cite{qu2024semantic}. 
Chunks retain relational metadata:
\begin{enumerate}
    \item Document ID – the original file identifier
    \item Chunk ID – sequence index within the document \\(e.g., doc1\_chunk1, doc1\_chunk2, …)
\end{enumerate}
This hierarchy ensures traceability and enables structured retrieval and analysis. Following the above approach, we obtain a total of $58144$ text chunks. With the corpus preprocessed and segmented into manageable chunks, the dataset is now prepared for topic modeling and label generation, which are detailed in Section~\ref{Tab:topic_modeling}, including their development and evolution.

\section{Topic Modeling and Label Generation}\label{Tab:topic_modeling}
In this section, we describe the topic modeling process and the subsequent generation of fine-grained, interpretable topic labels.
\subsection{Fine-Grained Topic Modeling}
We employ BERTopic~\cite{grootendorst2022bertopic}, a neural topic modeling technique that integrates transformer-based embeddings, density-based clustering, and class-based term frequency–inverse document frequency (c-TF-IDF) to extract semantically coherent and interpretable topics from unstructured text corpora. The modeling pipeline begins by generating sentence-level embeddings using a pre-trained language model~\cite{chen2024bge}, specifically the BAAI-bge-small-en\footnote{https://huggingface.co/BAAI/bge-small-en}) model. These embeddings capture fine-grained semantic relationships among textual units in a high-dimensional space.

To facilitate clustering, dimensionality reduction is performed using UMAP~\cite{mcinnes2018umap} (Uniform Manifold Approximation and Projection), which projects embeddings into a low-dimensional manifold while preserving local topological structure. Clusters are then identified using HDBSCAN~\cite{mcinnes2017accelerated} (Hierarchical Density-Based Spatial Clustering of Applications with Noise), an unsupervised algorithm well-suited for high-dimensional and variably dense data distributions typical of natural language.

Once clusters are formed, representative topic keywords are extracted via class-based TF-IDF. Unlike standard TF-IDF, which measures term relevance within individual documents, c-TF-IDF computes term importance over the concatenation of documents within each cluster. This formulation emphasizes terms that are frequent within a cluster yet infrequent across others, thereby yielding discriminative and interpretable topic descriptors. Formally, the c-TF-IDF value for a term t in class c is
calculated as:
\begin{equation}
    c-TF-IDF_{t,c} = \left(\frac{f_{t,c}}{ \sum_{\acute{t}\in V}f_{\acute{t},c}} \right) \cdot\log\left( \frac{N}{n_{t}} \right)
\end{equation}    
Where $f_{t,c}$ is the frequency of term $t$ in class $c$, $V$ is the vocabulary, $N$ is the total number of classes, and $n_t$ is the number of clusters in which term $t$ appears. $\sum_{\acute{t}\in V}f_{\acute{t},c}$ is the total number of all term occurrences in class/topic $c$
This formulation emphasizes words that are frequent within a cluster but rare across others, yielding discriminative topic keywords.

Following clustering, coherent groups of documents are identified, each characterized by a set of top-ranked keywords. These keywords are then input to a zero-shot topic labeling framework powered by language models.
\subsection{Topic Label Generation using Language Models} This section presents the configuration and infrastructure used to implement and evaluate the proposed topic discovery and semantic retrieval pipeline (Section~\ref{Tab:retrieval}). 
\subsubsection{Language models configuration}To generate interpretable topic labels after unsupervised topic modeling, we employed two pretrained language models (Flan and Mistral) under a zero-shot prompting paradigm~\cite{piper2025evaluating, kozlowski2024generative}:\\
\textbf{Flan-T5-Small.} We use a compact encoder-decoder model with 77 million parameters, chosen for its efficiency and adaptability in zero-shot settings. The model runs via Hugging Face Transformers\footnote{https://huggingface.co/google/flan-t5-small} and is executed in safetensors format (F32).\\
\textbf{Quantized Mistral-7B.} We use OpenHermes-2.5-Mistral-7B-GGUF, a powerful decoder-only model fine-tuned for conversational and summarization tasks. A quantized version is employed to reduce memory and computational overhead while maintaining output quality. The model is accessed via the Hugging Face Model Hub\footnote{https://huggingface.co/TheBloke/OpenHermes-2.5-Mistral-7B-GGUF}.


\subsubsection{Topic Labeling} 
\begin{figure*}[t]
  \centering
\includegraphics[width=0.99\textwidth]{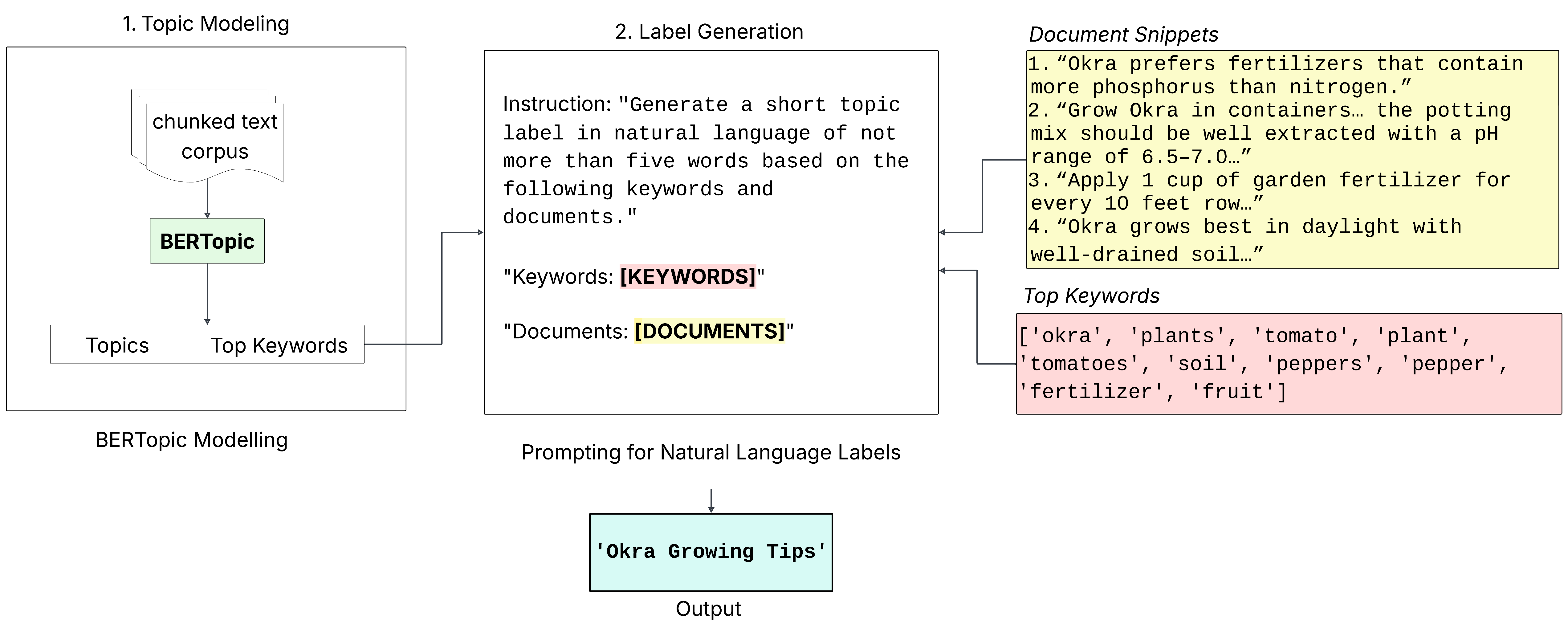}
  \caption{Workflow of topic label generation.}
  \label{fig:label_workflow}
\end{figure*}

The label generation process involves extracting topic keywords, generating prompts, and using zero-shot language models to produce interpretable labels. Figure~\ref{fig:label_workflow} illustrates this workflow, detailing the input components, language model interaction, and final label output. To formalize the topic-to-label process, we first define the output of the topic modeling step as follows:
Let the topic modeling step yield a set of $K$ topics =${T_1,T_2,…,T_K}$, where each topic $T_k$ is represented by:
\begin{itemize}
    \item a set of top-ranked keywords $w_k =\left\{ w_{k,1},…,w_{k,n}\right\}$, OR
    \item a set of representative documents $d_k=\left\{ d_{k,1},…,d_{k,m}\right\}$ (optionally).
\end{itemize}
 We define a language model $\mathcal{L}$ that takes as input a structured prompt $P_k$ derived from topic $T_k$, and returns a generated label $L_k$ for that topic:
\begin{equation}
     L_k  = \mathcal{L}\left( P_k\left( w_k, d_k, I \right) \right).
\end{equation}
Where $P_k(\cdot )$ is the prompt template used for topic $T_k$, $w_k$ are the keywords for topic $k$, $d_k$ are representative snippets, $I$  the instructional prompt constraining the LLM to generate short, domain-relevant, and human-readable topic labels.\\
Putting it all together, we define the Prompt-to-Label function as:
\textbf{}
\begin{equation}
    L_k = \mathcal{L}\left( I\parallel Keywords:w_k\parallel Snippets: d_k \right)
\end{equation}, where $\parallel$ denotes prompt concatenation.\\
Following this step, we obtain $85$ topic labels (Table~\ref{tab:topics}).
\subsubsection{Experimental setup}
To enable semantic retrieval (Section~\ref{Tab:retrieval}), we generate 384-dimensional embeddings using the all-MiniLM-L6-v2\footnote{https://huggingface.co/sentence-transformers/all-MiniLM-L6-v2} model from Sentence-BERT and index them in ChromaDB\footnote{https://pypi.org/project/chromadb/}, a lightweight vector database optimized for fast, large-scale retrieval. Experiments are conducted on Kaggle Notebooks with dual NVIDIA T4 GPUs.
\begin{table}
  \caption{Ten most frequently occurring topics in corpus, where F represents frequency.}
  \label{tab:topics}
  \begin{tabular}{| p{6cm}| p{0.8cm}|}
    \toprule
     \textbf{Topic Label Generated by LMs} &\textbf{F}\\
    \midrule
    Food Security and Climate Change &4450\\
    Sauce, Recipe, Chicken, Cook,  and Ingredients &3467\\
Agriculture and Crop Management &3097\\
Tractor Engines and Models &2373\\
Canada Agriculture Investment &2362\\
Market News &2325\\
Dairy Farming and Livestock Management &2297\\
Plant Nutrition and Fertilization &2274\\
USDA Climate Agriculture Programs &1697\\
UK-New Zealand Trade Deal &1406\\
 \bottomrule
\end{tabular}
\end{table}\\
\subsubsection{Evaluation of Generated Labels} We evaluate the generated topic labels using multiple criteria, including bias and semantic consistency, factuality and grounding, document coverage, hallucination analysis, and a qualitative human assessment.

\textbf{Bias and Semantic Evaluation.} 
To assess the semantic alignment and representational fidelity of the generated topic labels, we conducted a comparative evaluation using Cosine Similarity and BERTScore metrics. The Cosine Similarity was computed between sentence embeddings of the model-generated labels and the original topic representations using the all-MiniLM-L6-v2 model. Additionally, BERTScore, which leverages contextual embeddings from a pre-trained BERT model to capture deep semantic equivalence, was used to further quantify alignment.

\begin{table}[t]
  \caption{Bias and semantic evaluation of the generated labels. }
  \label{tab:exp2}
  \centering
  \begin{tabular}{|p{2.5cm}|p{2.7cm}|p{1.8cm}|}
    \hline
    \textbf{Language Model} & \textbf{Average Cosine Similarity} & \textbf{BERTScore (F1)} \\
    \hline
   Flan-T5-small &0.640 &0.814\\
   \hline
    Mistral-7B &\textbf{0.699} & \textbf{0.825}\\
    \bottomrule
  \end{tabular}
\end{table}
As shown in Table~\ref{tab:exp2}, both models exhibit a high degree of semantic similarity with the reference topic descriptions. Mistral-7B consistently outperforms Flan-T5, achieving a higher Cosine Similarity (0.699 vs. 0.640) and a superior BERTScore (0.825 vs. 0.814). These results suggest that Mistral-7B generates topic labels that are more semantically faithful to the original representations, potentially due to its larger parameter count and stronger contextual understanding.\\
\textbf{Factuality and Grounding Evaluation. } 
To assess the quality and faithfulness of generated topic labels, we evaluated models along two dimensions: \textit{Factuality}, which measures whether the topic label accurately represents the underlying document content, and \textit{Document Coverage}, which quantifies how well the retrieved chunks cover the original topic’s document distribution. We define the \textbf{Factuality Score} by first tokenizing the generated label into individual words, then calculating the proportion of these words that appear in the corresponding representative documents.
\begin{equation}
\resizebox{\linewidth}{!}{$
\text{Factuality Score} =
\frac{\text{No. of labeled words found in topic documents}}
{\text{Total labeled words}}
$}
\end{equation}
\textbf{Document Coverage Score (Per topic). } For each label word, we check in how many of the top documents that word appears.
Then report the proportion of top documents containing at least one label word.

As shown in Table~\ref{tab:exp3}, Flan-T5 achieved a slightly higher factuality score (0.90) compared to Mistral-7B (0.87), suggesting that its topic labels were more aligned with the factual content of the source documents. However, Mistral-7B excelled in Document Coverage, achieving near-perfect grounding (0.99), slightly outperforming Flan-T5 (0.97).
These results highlight a trade-off between precision and breadth: Flan-T5 offers stronger factual grounding, while Mistral-7B provides broader topical coverage across the corpus. 
\begin{table}[t]
  \caption{Factuality and document coverage scores for topic labels.}
  \label{tab:exp3}
  \centering
  \begin{tabular}{|p{2.5cm}|p{2cm}|p{2cm}|}
    \hline
    \textbf{Language Model} & \textbf{Factuality } & \textbf{Document Coverage} \\
    \hline
   Flan-T5-small & \textbf{0.90} &0.97\\
   \hline
    Mistral-7B &0.0.87 &\textbf{0.99}\\
    \bottomrule
  \end{tabular}
\end{table}\\
\textbf{Hallucination Analysis. } To flag hallucinated outputs, we extract cases where:
$\text{ Cosine similarity} > 0.7 \rightarrow$ the generated label sounds plausible or semantically aligns with expected themes.
$\text{Factuality} < 0.3 \rightarrow $ the label is not actually grounded in the underlying data (i.e., it misrepresents the topic cluster).
These are instances where the model appears confident and coherent, yet introduces content not supported by the data—a classic case of LLM hallucination.
This method helps disentangle fluency from faithfulness. Language models often generate fluent outputs, and high semantic similarity can give the illusion of correctness. By combining it with a low factuality threshold, we catch cases where the model's output might "sound right" but is actually factually misleading or imagined.
Such hallucinations are especially important to detect in scientific or domain-specific contexts like agriculture, where false confidence in a topic label can mislead researchers or downstream applications.\\
\textbf{Human Evaluation.} To further assess the quality of topic labels generated by the two models—Flan-T5 and a quantized version of Mistral-7B—we conducted a human evaluation on a randomly sampled set of 10 topics. For each topic, annotators reviewed the representative documents and the corresponding labels produced by both models. Evaluation was carried out across three criteria: Relevance (the degree to which the label reflected the content of the documents), Abstraction (the ability of the label to generalize beyond literal phrases), and Fluency (the naturalness and readability of the label). Each criterion was scored on a 5-point Likert scale.
\begin{table}[t]
  \caption{Human Evaluation of the generated topic labels}
  \label{tab:exp4}
  \centering
  \begin{tabular}{|p{2.2cm}|p{1.5cm}|p{1.7cm}|p{1.5cm}|}
    \hline
    \textbf{Language Model} & \textbf{Relevance} & \textbf{Abstraction} &\textbf{Fluency}  \\
    \hline
   Flan-T5-small &4.1/5 &3.4/5 &3.6/5\\
   \hline
    Mistral-7B & \textbf{4.3/5} &\textbf{4.2/5} &\textbf{4.4/5}\\
    \bottomrule
  \end{tabular}
\end{table}
The results in Table~\ref{tab:exp4} indicate that both models generated relevant and grounded labels, avoiding factual hallucinations. However, Mistral-7B consistently outperformed Flan-T5 in abstraction and fluency. Mistral-7B’s labels were more concise, human-like, and semantically generalized. In contrast, Flan-T5 tended to produce more literal or extractive outputs, often reusing sentences or phrases from the source text. These findings suggest that Mistral-7B is better suited for generating high-level topic descriptors, particularly in applications where interpretability and natural phrasing are prioritized.

\textit{As the label generation performance of both models were found to be comparable, we selected Mistral-7B for the end-to-end system implementation.} Accordingly, this model is used in the next section (Section~\ref{Tab:retrieval}) for further evaluation in the semantic retrieval pipeline.


\section{Semantic Retrieval and Evaluation} \label{Tab:retrieval}
Building on the extracted topic representations, we implement a topic-guided semantic retrieval pipeline using Mistral-7B that enables interactive search and summarization over document embeddings. Users can query using the generated natural language topic labels to retrieve semantically aligned content. To complement this, a bias and semantic evaluation module assesses results for representational balance and coherence, offering insights into the fairness and quality of topic coverage.
\subsubsection{Topic Guided Semantic Retrieval}The system matches high-dimensional document embeddings with semantically similar inputs, using both topic keywords and user queries. This allows users to navigate the corpus through interpretable topics or refine searches with natural language, guided by topic-level semantics. The following provides a mathematical formulation of the process:\\
\textbf{Mathematical Formulation.} Let the corpus $\mathcal{D} = \left\{ d_1, d_2, ...., d_N \right\}$ be a set of documents, where each document $d_i$ is segmented into a set of smaller, semantically coherent chunks $\left\{ c_{i1},c_{i2},....c_{1M}\right\}$. \\Each chunk $c_{ij}$ is mapped to a dense vector representation $v_{ij}\in \mathrm{R}$ using a pre-trained sentence embedding model $f: text \to \mathrm{R}^d$. All chunk vectors $v_{ij}$ are stored in a vector database $\mathcal{V}$, along with their corresponding metadata, including chunk ID: $id_{ij}$ and document ID: $id_i$, enabling traceability back to the original document structure. For each extracted topic label $t_k$, when the user passes it as a query, it is first converted into a dense vector representation $q_k = f(t_k)$ using the same sentence embedding model. This vectorized query is then used to retrieve semantically relevant chunks from the database. To achieve this, a \textbf{top-k} similarity search is performed over the stored vectors in the vector database:
\begin{equation}
\mathrm{R}_k = \text{Top-k}\left\{ \cos(q_k,v_{ij}))| \forall v_{ij}\in \mathcal{V} \right\},
\end{equation}
where $\cos(a,b) = \frac{a\cdot b}{\left\| a \right\|\left\| b \right\|}$ denotes the cosine similarity between the topic query and each stored chunk.
\subsubsection{Human Evaluation of Retrieved Chunks}
To evaluate the quality of retrieved content, we conducted a human evaluation\footnote{https://www.evidentlyai.com/llm-guide/llm-evaluation-metrics} assessing semantic alignment, informativeness, and coherence of document chunks retrieved via topic-based vector search, using the generated topic labels as natural language queries~\cite{isonuma2024comprehensive}. Three human annotators with backgrounds in agriculture and information retrieval participate in the evaluation. Each annotator independently assesses the retrieved chunks without access to the original documents or retrieval method. \\
\textbf{Evaluation Criteria.} Each retrieved chunk was assessed based on four dimensions, inspired by prior work in semantic retrieval and human-centered IR evaluation~\cite{taherdoost2019best,elangovan2024considers,sperrle2021survey}:
\begin{itemize}
    \item \textit{Relevance.} The degree to which the chunk addresses the topic label semantically and contextually.
    \item \textit{Specificity.} The extent to which the chunk provides detailed and non-generic information.
    \item \textit{Coherence.} The grammatical and structural completeness of the chunk; whether it forms a well-formed unit of meaning.
    \item \textit{Usefulness.} The perceived utility of the chunk in supporting user understanding or exploration of the topic.
\end{itemize}
 Each dimension was rated on a 3-point Likert scale 0 = Poor, 1 = Moderate, and 2 = Strong. Each chunk is scored out of a maximum of 0 points. Inter-annotator agreement is measured using Fleiss’ Kappa~\cite{fleiss1973equivalence}.
\begin{table}
  \caption{Evaluation of retrieved chunks using Mistral-7B across four dimensions reflecting semantic alignment and summarization usefulness. }
  \label{tab:semantic_eval}
  \centering
  \begin{tabular}{| p{4cm}|p{3cm}|}
    \toprule
     \textbf{Evaluation Dimension} &\textbf{Mean Score (0-2)}\\
    \midrule
    Relevance &1.85\\
    Specificity &1.42\\
    Coherence &1.68\\
    Usefulness &1.63\\
    \hline
    \textbf{Overall Mean} & \textbf{6.3/9.0}\\
 \bottomrule
\end{tabular}
\end{table}\\
As shown in Table~\ref{tab:semantic_eval} relevance score is high, showing strong topical alignment between retrieved chunks and query labels. Specificity scores lower, indicating that some chunks, while relevant, provide only general or surface-level information.
Well-formed, specific topic labels produce more coherent and informative results, while broad or abstract labels lead to partial or ambiguous retrieval.
This evaluation shows that the system retrieves content that is both relevant and understandable to human readers.
\section{ Deployment Considerations and User Scenarios}
\begin{figure}[h]
  \centering
\includegraphics[width=0.99\linewidth]{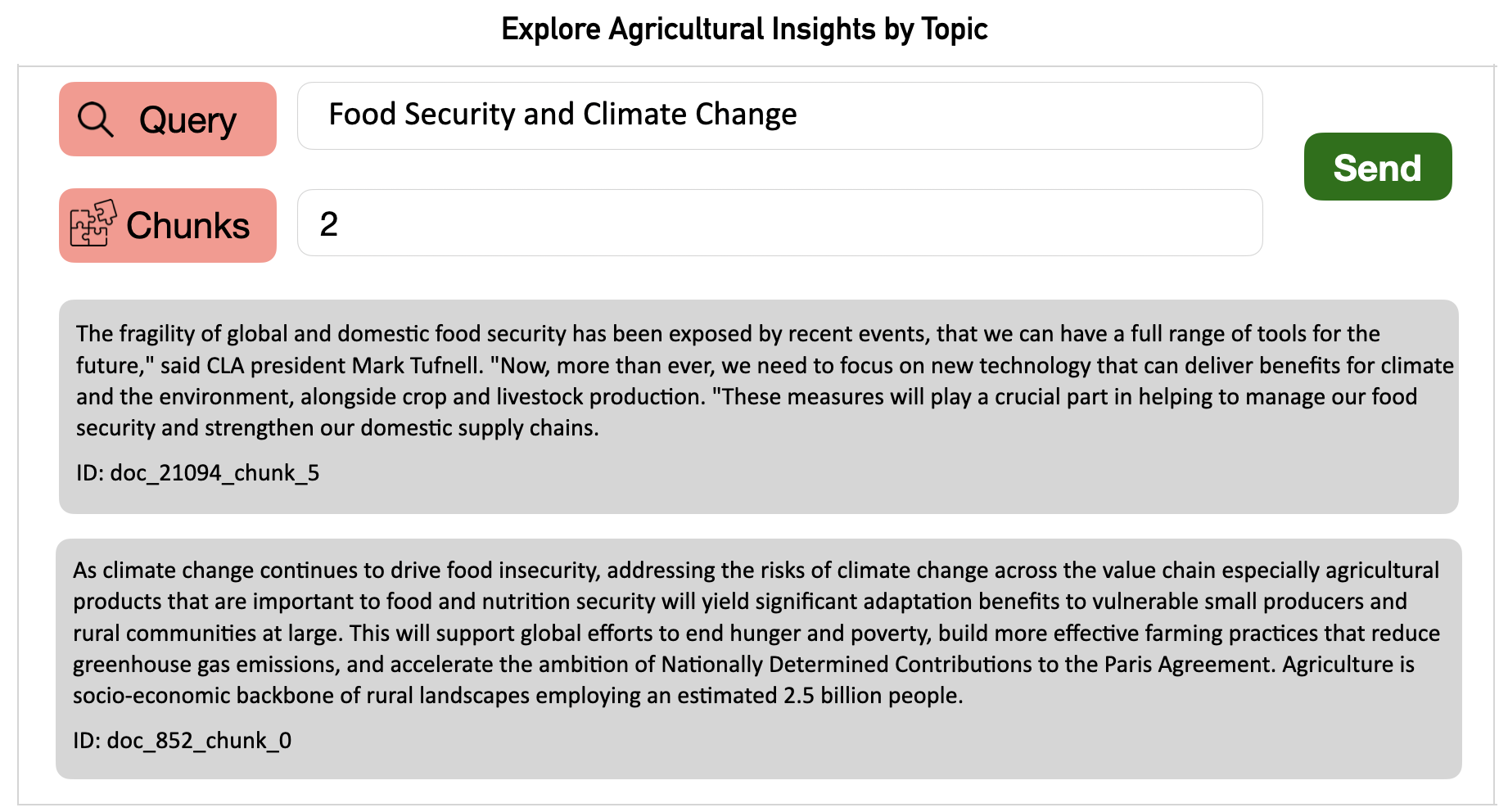}
  \caption{A snapshot of AgriLens system usage.}
  \label{fig:usage_agrilens_system}
\end{figure}

To support real-world exploration of topic-specific agricultural insights, we developed a lightweight user interface that enables end-users to query and retrieve semantically relevant document segments. Figure~\ref{fig:usage_agrilens_system} shows the deployed prototype interface, where users can input a topic query (e.g., Food Security and Climate Change) and specify the number of supporting chunks to retrieve. The interface then displays the most semantically aligned document chunks returned by the retrieval system.
In practice, such a system can support agricultural policy design, climate impact studies, and sector-specific reporting by surfacing thematically linked insights with minimal manual filtering.

\section{Conclusion}
We proposed a comprehensive framework that combines neural topic modeling, zero-shot topic labeling, and topic-guided semantic retrieval. The system produces coherent topic structures, assigns interpretable labels using language models, and supports retrieval aligned with user queries. By integrating a bias evaluation module, the framework also facilitates responsible deployment and analysis of large textual datasets. Our results demonstrate strong document coverage, meaningful topic-label alignment, and effective semantic search—all without requiring manual annotation. This approach offers a scalable solution for exploring, summarizing, and understanding unstructured text across domains.
\section*{Acknowledgement}
We thank Brett Drury and Omobolaji Adelowo, the creators of the ENAGRINEWS dataset, for making the data available and supporting research in agricultural text analysis.

\end{document}